\documentstyle[preprint,aps,epsf]{revtex}

\def\singlespace {\smallskipamount=3.75pt plus1pt minus1pt
                  \medskipamount=7.5pt plus2pt minus2pt
                  \bigskipamount=15pt plus4pt minus4pt
                  \normalbaselineskip=12pt plus0pt minus0pt
                  \normallineskip=1pt
                  \normallineskiplimit=0pt
                  \jot=3.75pt
                  {\def\smallskip {\vskip\smallskipamount}}
                  {\def\medskip   {\vskip\medskipamount}}
                  {\def\bigskip   {\vskip\bigskipamount}}
                  {\setbox\strutbox=\hbox{\vrule
                    height10.5pt depth4.5pt width 0pt}}
                  \parskip 7.5pt
                  \normalbaselines}
\def\middlespace {\smallskipamount=5.625pt plus1.5pt minus1.5pt
                  \medskipamount=11.25pt plus3pt minus3pt
                  \bigskipamount=22.5pt plus6pt minus6pt
                  \normalbaselineskip=22.5pt plus0pt minus0pt
                  \normallineskip=1pt
                  \normallineskiplimit=0pt
                  \jot=5.625pt
                  {\def\smallskip {\vskip\smallskipamount}}
                  {\def\medskip   {\vskip\medskipamount}}
                  {\def\bigskip   {\vskip\bigskipamount}}
                  {\setbox\strutbox=\hbox{\vrule
                    height15.75pt depth6.75pt width 0pt}}
                  \parskip 11.25pt
                  \normalbaselines}
\def\doublespace {\smallskipamount=7.5pt plus2pt minus2pt
                  \medskipamount=15pt plus4pt minus4pt
                  \bigskipamount=30pt plus8pt minus8pt
                  \normalbaselineskip=30pt plus0pt minus0pt
                  \normallineskip=2pt
                  \normallineskiplimit=0pt
                  \jot=7.5pt
                  {\def\smallskip {\vskip\smallskipamount}}
                  {\def\medskip   {\vskip\medskipamount}}
                  {\def\bigskip   {\vskip\bigskipamount}}
                  {\setbox\strutbox=\hbox{\vrule
                    height21.0pt depth9.0pt width 0pt}}
                  \parskip 15.0pt
                  \normalbaselines}
\def\al{\alpha}

\def\de{\delta}
\def\ep{\epsilon}

\def\th{\theta}

\def\si{\sigma}

\def\ph{\phi}

\def\Ph{\Phi}

\def\cF{{\cal F}}
\def\cL{{\cal L}}

\def\frac#1#2{\textstyle{{{#1} \over {#2}}}}

\def\prt{\partial}

\def\half{{\textstyle{1\over 2}}}
\def\lsim{\mathrel{\rlap{\lower4pt\hbox{\hskip1pt$\sim$}}
    \raise1pt\hbox{$<$}}}
\def\gsim{\mathrel{\rlap{\lower4pt\hbox{\hskip1pt$\sim$}}
    \raise1pt\hbox{$>$}}}
\def\ol#1{\overline{#1}}

\newcommand{\beq}{\begin{equation}}
\newcommand{\eeq}{\end{equation}}
\newcommand{\bea}{\begin{eqnarray}}
\newcommand{\eea}{\end{eqnarray}}
\newcommand{\rf}[1]{(\ref{#1})}

\tightenlines
\begin{document}
\preprint{
\hfill$\vcenter{\hbox{\bf IUHET-457} \hbox{August
             2003}}$  }

\title{\vspace*{.75in}
Superfield Realizations of Lorentz and CPT Violation}

\author{M. S. Berger
\footnote{Electronic address:
berger@indiana.edu}}

\address{
Physics Department, Indiana University, Bloomington, IN 47405, USA}

\maketitle

\thispagestyle{empty}

\begin{abstract}
Superfield realizations of Lorentz-violating extensions of the Wess-Zumino 
model are presented. These models retain supersymmetry but include 
terms that explicitly break the Lorentz symmetry. The models can be 
understood as arising from superspace transformations that are modifications
of the familiar one in the Lorentz-symmetric case.
\end{abstract}

\pacs{11.30.Pb, 11.30.Cp, 11.30.Er}

\newpage

\section{Introduction}
Spacetime symmetries have played an important role in formulation of  
theories of fundamental physics for the last hundred years. The special and
general theories of relativity are founded on underlying spacetime symmetries,
and the most popular speculations about the possible advances in physics 
involve further spacetime symmetries such as 
supersymmetry\cite{sw}. 
If supersymmetry does in fact describe nature, then it is clear from
experimental observation that this
is a symmetry that must be broken. 

Since spacetime symmetries are central to fundamental particle physics, it is 
important to consider the consequences of all possible ways of 
breaking them. Broken
supersymmetry has been extensively studied because of the experimental 
necessity of splitting the masses of the observed particles from their 
supersymmetric partners. The Lorentz symmetry requires that their be
no preferred direction in space and no preferred frame. 
Whether this symmetry is exact or 
broken is a question for experiment. 

It is usually considered desirable for a broken symmetry to arise spontaneously
since then certain properties of the theory that result from the underlying 
symmetry are retained. Furthermore, it is expected that any symmetry that is 
approximately valid must have some fundamental realization, and any breaking
should 
arise spontaneously. Presumably the explicit supersymmetry breaking terms 
that are added to the models used in phenomenological supersymmetric theories
such as the Minimal Supersymmetric Standard Model arise in a more fundamental
theory from some spontaneous breaking of supersymmetry. Nevertheless, these 
low-energy supersymmetric theories can be viewed as effective theories.

If one believes in electroweak-scale 
supersymmetry, one has to accept that a spacetime symmetry is broken. 
From the point of view of available experiment, this
breaking seems to be a very large effect; in fact, the supersymmetry breaking
is so large that no superpartners have even been found. However, compared to 
the fundamental Planck scale, the supersymmetry breaking scale is very small, 
and supersymmetry is more appropriately viewed as an approximate symmetry. 
Indeed major effort has gone into trying to understand why the breaking of 
supersymmetry is so small compared to the Planck scale.

Recent 
articles\cite{Berger:2001rm,Berger:2002km,Belich:2003fa}
have studied the possibility of introducing Lorentz-violation into 
supersymmetric theories.
Most theories incorporate the Poincar\'e symmetry at the outset. This 
assumption is well-justified on the basis that no violations have ever been 
observed experimentally. However, in light of the argument just made that 
supersymmetry is a approximate spacetime symmetry, it seems appropriate to 
consider the possibility that there is violation of the other spacetime
symmetries at some level. 
Supersymmetry can be broken within the context of local 
field theory. Lorentz and CPT
violation might arise from nonlocal interactions in a more fundamental theory. 
One approach 
is to use a field theory treatment of Lorentz and CPT-violation that
incorporates their effects by adding explicit terms to a symmetric Lagrangian
and the resulting field theories should be regarded as effective theories
only. Problems with microcausality are addressed in the underlying fundamental 
theory at the energy scales at which the effective theory breaks
down\cite{Kostelecky:2000mm}. The experimental implications of Lorentz and 
CPT-violation have been explored extensively in recent years\cite{cpt01}.

In Ref.~\cite{Berger:2001rm} we examined the possibility that one could 
construct a Lagrangian that respects a supersymmetry algebra, but that has 
terms that explicitly violate the Lorentz symmetry. Since these models contain
Lorentz-violation, they fall outside the usual classification of supersymmetry
algebras\cite{Haag:1974qh}. The conventional 
supersymmetry algebra is given by a transformation (involving a two-component
Weyl spinor $Q$) between bosons and fermions
that upon anticommutation yields the translation operator:
\bea
\left [Q,P_\mu \right ] &=&0\nonumber \\
\left \{Q,\overline{Q}\right \}&=&2\si ^\mu P_\mu \;.
\label{susy}
\eea
It was shown that there are indeed simple extensions of the Wess-Zumino 
model\cite{Wess:tw} 
that respect a similar algebra, but that have extra terms that explicitly 
violate the Lorentz symmetry characterized by the generators of boosts and 
rotations, $M_{\mu \nu}$,
\bea
\left [ P_\mu,P_\nu \right ] &=&0\nonumber \\
\left [ P_\mu,M_{\rho \sigma}\right ]&=&i(\eta _{\mu \rho}P_\sigma
-\eta _{\mu \sigma}P_\rho)\nonumber \\
\left [ M_{\mu \nu}, M_{\rho \sigma}\right ] &=& i(\eta _{\nu \rho}
M_{\mu \sigma}-\eta _{\nu \sigma}M_{\mu \rho}
-\eta _{\mu \rho}M_{\nu \sigma}
+\eta _{\mu \sigma}M_{\nu \rho})\;,
\label{poincare}
\eea
The commutation relations
in Eqn.~\rf{poincare} form the Poincar\'e algebra, and when
the supersymmetric generators are included as in Eqn.~\rf{susy}, the result is
the superPoincar\'e algebra. Lorentz-violating models will not respect all 
the commutation relations involving $M_{\mu\nu}$.

Two extensions to the supersymmetric 
Wess-Zumino model were presented which contain explicit terms contain the 
Lorentz-violation. One of these models preserves CPT whereas the other is 
CPT-violating. A superspace formulation for the CPT preserving theory was 
already presented in Ref.~\cite{Berger:2001rm}. In this paper we extend 
the superspace formulation to the CPT-violating model. We show that the 
Wess-Zumino model and its two extensions all admit a description in terms of 
transformations on superspace. In addition, we present
the two extended models in an alternative form involving two-component Weyl 
spinors as opposed to the four-component Majorana spinors used in 
Ref.~\cite{Berger:2001rm}.

The usual Wess-Zumino Lagrangian is elegantly derived in the framework of 
superspace\cite{Salam:1974yz}. 
A superfield $\Ph(x,\th,\bar\th)$ is a function
of the commuting spacetime coordinates $x^\mu$ and of four anticommuting 
coordinates $\th ^\al$ and $\bar\th _{\dot\al}$ which form two-component Weyl 
spinors.
A chiral superfield is a function of 
$y^\mu =x^\mu+i\th \si ^\mu \bar \th$ and
$\th$, i.e.
\bea 
\Ph(x,\th,\bar\th)&=&\ph(y)+\sqrt{2}\th \psi(y)+(\th \th)\cF(y)\;, \nonumber \\
&=&\ph(x)+i\th \si ^\mu \bar\th\prt _\mu \ph(x)
-{1\over 4}(\th \th)(\bar\th \bar\th)\Box \ph(x)\nonumber \\
&&+\sqrt{2}\th \psi(x)+i\sqrt{2}\th \si ^\mu \bar\th \th\prt _\mu \psi(x)
+(\th \th)\cF(x)
\eea
where one can define the usual real components of the complex scalar components
as
\bea
&&\ph = \frac 1 {\sqrt 2} (A + i B),
\quad
\cF = \frac 1 {\sqrt 2} (F - i G).
\eea
The conjugate superfield is 
\bea 
\Ph^*(x,\th,\bar\th)&=&\ph^*(z)+\sqrt{2}\bar\th \bar\psi(z)
+(\bar\th \bar\th)\cF^*(z)
\;, \nonumber \\
&=&\ph^*(x)-i\th \si ^\mu \bar\th\prt _\mu \ph^*(x)
-{1\over 4}(\th \th)(\bar\th \bar\th)\Box \ph^*(x)\nonumber \\
&&+\sqrt{2}\bar\th \bar\psi(x)+i\sqrt{2}\th \si ^\mu \bar\th \bar\th\prt _\mu 
\bar\psi(x)
+(\bar\th \bar\th)\cF^*(x)
\eea
where $z^\mu=x^\mu -i\th \si ^\mu \bar \th=y^{\mu *}$.
We have taken the opportunity to reexpress the superfields in terms
of Weyl spinors as opposed to the use of Majorana spinors in 
Ref.~\cite{Berger:2001rm}.

The Lagrangian can be derived from the superspace integral
\bea
\int d^4\th \Ph^*\Ph + \int d^2\th \left [ 
{1\over 2}m\Ph^2 +{1\over 3}g\Ph^3 
+h.c.\right ]
\label{superspace}
\eea
where the superspace integrals elegantly project out the 
$(\th \th)(\bar \th \bar\th)$ component of the $\Ph^* \Ph$ 
superfield in the first term, and the $\th \th$ component in the second term.
The result is the Wess-Zumino Lagrangian,
\bea
\cL&=&\prt _\mu \phi^* \prt ^\mu \ph +{i\over 2}[(\prt _\mu \psi) \si ^\mu 
\bar\psi+(\prt _\mu\bar\psi)\bar\si ^\mu \psi]
+\cF^*\cF \nonumber \\
&&+m\left [\ph \cF + \ph ^*\cF^* -\half\psi\psi -\half\bar\psi\bar\psi\right ] 
\nonumber \\
&&+g\left [\ph^2\cF+\ph^{*2}\cF^*-\ph (\psi\psi)-\ph^*(\bar\psi\bar\psi)
\right ]\;.
\label{wz}
\eea
The action $\de_S \Ph (x,\th,\bar\th) = - i (\ep Q +\bar\ep \bar Q)
\Ph(x,\th,\bar\th)$ of the supersymmetry generators $Q$ and $\bar Q$
\bea
&&\delta _S\Ph(x,\th,\bar\th)=\left [\ep ^\al \prt _\al
+\bar\ep _{\dot\al} \bar\prt^{\dot\al}+i\th \si ^\mu\bar\ep \prt _\mu
-i\ep \si ^\mu\bar\th \prt _\mu\right ]\Ph(x,\th,\bar\th)\;,
\eea
transforms the Lagrangian into itself plus a total derivative.

As shown in Ref.~\cite{Berger:2001rm}, Lorentz-violation can be introduced
into the Wess-Zumino Lagrangian via the substitution 
$\prt _\mu \to \prt _\mu +k_{\mu\nu}\prt ^\nu$,
\bea
\cL_{\rm Lorentz}&=&
(\prt _\mu +k_{\mu\nu}\prt ^\nu)\phi^* (\prt ^\mu +k^{\mu\rho}\prt _\rho)
\ph \nonumber \\
&&+{i\over 2}[((\prt _\mu +k_{\mu\nu}\prt ^\nu)\psi) \si ^\mu 
\bar\psi+((\prt _\mu+k_{\mu\nu}\prt ^\nu)\bar\psi)\bar\si ^\mu \psi]
+\cF^*\cF \nonumber \\
&&+m\left [\ph \cF + \ph ^*\cF^* -\half\psi\psi -\half\bar\psi\bar\psi\right ] 
\nonumber \\
&&+g\left [\ph^2\cF+\ph^{*2}\cF^*-\ph (\psi\psi)-\ph^*(\bar\psi\bar\psi)
\right ]\;.
\label{Lorentz}
\eea
In this equation,
$k_{\mu\nu}$ is a real, symmetric, traceless, 
and dimensionless coefficient 
determining the magnitude of Lorentz-violation.
The coefficient $k_{\mu\nu}$
transforms as a 2-tensor under observer Lorentz transformations
but as a scalar under particle Lorentz transformations
\cite{Colladay:1996iz,Colladay:1998fq}. 

An extension of the Wess-Zumino model that violates CPT in addition to 
containing Lorentz-violation was introduced in 
Ref.~\cite{Berger:2001rm}. The Lagrangian for the model is
\bea
\cL_{\rm CPT}&=&\left [(\prt _\mu-ik_\mu) \phi^*\right ]
\left [ (\prt ^\mu +ik^\mu)\ph\right ] 
+{i\over 2}[((\prt _\mu +ik_\mu)\psi) \si ^\mu 
\bar\psi+((\prt _\mu-ik_\mu)\bar\psi)\bar\si ^\mu \psi]
+\cF^*\cF\;.
\label{CPT}
\eea
Here the Lorentz and CPT-violation is controlled by $k_\mu$, which is a real 
coefficient of mass dimension one transform as a four-vector under observer 
Lorentz transformations but is unaffected by particle Lorentz 
transformations\cite{Colladay:1996iz,Colladay:1998fq}. Unlike the coefficient
$k_{\mu \nu}$, the quantity $k_\mu$ as an odd number of four-indices so it
violates CPT. It has been shown on quite general grounds
that CPT-violation implies Lorentz 
violation\cite{Greenberg:2002uu,Greenberg:2003ks}.
The Lagrangian for the model with the CPT-violating coefficient $k_\mu$ 
can be obtained from the kinetic part of the Wess-Zumino Lagrangian in 
Eqn.~\rf{wz} with the appropriate substitutions 
$\prt _\mu \to \prt _\mu \pm ik_\mu$.

\section{Modified Superfields}

The two Lorentz-violating models can be understood in the superspace 
formalism in a way that parallels that of the ordinary Wess-Zumino model.
Define superfields
\bea 
\Ph_y(x,\th,\bar\th)
&=&\Ph(x,\th,\bar\th;\prt _\mu \to \prt _\mu +k_{\mu\nu}\prt ^\nu)\nonumber \\
&=&\ph(x_+)+\sqrt{2}\th \psi(x_+)+(\th \th)F(x_+)\;, \nonumber \\
&=&\ph(x)+i\th \si ^\mu \bar\th(\prt _\mu +k_{\mu \nu}\prt ^\nu) \ph(x)
-{1\over 4}(\th \th)(\bar\th \bar\th)(\prt _\mu +k_{\mu \nu}\prt ^\nu)
(\prt ^\mu +k^{\mu \rho}\prt _\rho) \ph(x)\nonumber \\
&&+\sqrt{2}\th \psi(x)+i\sqrt{2}\th \si ^\mu \bar\th \th
(\prt _\mu +k_{\mu \nu}\prt ^\nu) \psi(x)
+(\th \th)F(x)\;,
\eea
and 
\bea 
\Ph^*_y(x,\th,\bar\th)
&=&\Ph^*(x,\th,\bar\th;\prt _\mu \to \prt _\mu +k_{\mu\nu}\prt ^\nu)
\nonumber \\
&=&\ph^*(x_-)+\sqrt{2}\bar\th \bar\psi(x_-)+(\bar\th \bar\th)F^*(x_-)
\;, \nonumber \\
&=&\ph^*(x)-i\th \si ^\mu \bar\th(\prt _\mu +k_{\mu \nu}\prt ^\nu) \ph^*(x)
-{1\over 4}(\th \th)(\bar\th \bar\th)(\prt _\mu +k_{\mu \nu}\prt ^\nu)
(\prt ^\mu +k^{\mu \rho}\prt _\rho) \ph^*(x)\nonumber \\
&&+\sqrt{2}\bar\th \bar\psi(x)+i\sqrt{2}\th \si ^\mu \bar\th \bar\th
(\prt _\mu +k_{\mu \nu}\prt ^\nu) 
\bar\psi(x)
+(\bar\th \bar\th)F^*(x)
\eea
where 
\bea
&&x_\pm^\mu =x^\mu \pm i\th \si ^\mu \bar\th \pm ik^{\mu \nu}\th 
\si_\nu\bar\th\;.,
\eea
are shifted coordinates that take the place of $y^\mu$ and $z^\mu$.
Under a CPT-transformation the chiral superfield $\Ph_y$ and the 
antichiral superfield $\Ph^*_y$ transform into themselves just as the usual
superfields $\Ph$ and $\Ph^*$ do. The Lagrangian in Eqn.~\rf{Lorentz} can 
be obtained by the same superspace integral in Eqn.~\rf{superspace} with 
the superfields $\Ph_y$ and $\Ph^*_y$ substituted in the place of 
$\Ph$ and $\Ph^*$ (see Eqn.~\rf{superspace2} below).

As argued in Ref.~\cite{Berger:2001rm}, it is clear that the Lagrangian in 
Eqn.~\rf{Lorentz} transforms into a total derivative 
under the supersymmetric transformation
\bea
&&\delta _S\Ph_y(x,\th,\bar\th)=\left [\ep ^\al \prt _\al
+\bar\ep _{\dot\al} \bar\prt^{\dot\al}+i\th \si ^\mu\bar\ep (\prt _\mu 
+k_{\mu\nu}\prt ^\nu)
-i\ep \si ^\mu\bar\th (\prt _\mu 
+k_{\mu\nu}\prt ^\nu)\right ]\Ph_y(x,\th,\bar\th)
\eea
since it is simply the usual supersymmetric transformation in Eqn.~\rf{susy}
with the substitution $\prt _\mu \to \prt _\mu +k_{\mu\nu}\prt ^\nu$.
The superalgebra generated by $Q$
and $P_\mu = i \prt_\mu$ is 
\bea
&&\left [P_\mu, Q \right ] =0,
\quad
\left \{Q,\ol{Q}\right \}
=2\si ^\mu P_\mu  + 2 k_{\mu\nu}\si^\mu P^\nu.
\label{superalg}
\eea

Consider now the CPT-violating model in Eqn~\rf{CPT}. 
Define modified superfields by the substitutions
\bea 
\Ph_k(x,\th,\bar\th)
&=&\Ph(x,\th,\bar\th;\prt _\mu \to \prt _\mu +ik_\mu)\nonumber \\
&=&\ph(x)+\th \si ^\mu \bar\th(i\prt _\mu -k_\mu)\ph(x)
+{1\over 4}(\th \th)(\bar\th \bar\th)(i\prt _\mu -k_\mu)(i\prt ^\mu -k^\mu) 
\ph(x)\nonumber \\
&&+\sqrt{2}\th \psi(x)+\sqrt{2}\th \si ^\mu \bar\th \th(i\prt ^\mu -k^\mu) 
\psi(x)
+(\th \th)F(x)\;,
\eea
and
\bea 
\Ph^*_k(x,\th,\bar\th)
&=&\Ph^*(x,\th,\bar\th;\prt _\mu \to \prt _\mu -ik_\mu)\nonumber \\
&=&\ph^*(x)-\th \si ^\mu \bar\th(i\prt _\mu +k_\mu)\ph^*(x)
+{1\over 4}(\th \th)(\bar\th \bar\th)(i\prt _\mu +k_\mu)(i\prt ^\mu +k^\mu) 
 \ph^*(x)\nonumber \\
&&+\sqrt{2}\bar\th \bar\psi(x)+\sqrt{2}\th \si ^\mu \bar\th \bar\th
(i\prt _\mu +k_\mu)
\bar\psi(x)
+(\bar\th \bar\th)F^*(x)\;.
\eea
The 
infinitesimal supersymmetry transformations acting on a superfield ${\cal S}$
\bea
\delta _S\Ph_k(x,\th,\bar\th)&=&\left [\ep ^\al \prt _\al
+\bar\ep _{\dot\al} \bar\prt^{\dot\al}+i\th \si ^\mu\bar\ep (\prt _\mu +ik_\mu)
-i\ep \si ^\mu\bar\th (\prt _\mu-ik_\mu)\right ]\Ph_k(x,\th,\bar\th)
\nonumber \\
&=&\left [\ep ^\al (\prt +k_\mu \si ^\mu\bar\th)_\al
+\bar\ep _{\dot\al} (\prt -k_\mu \th \si ^\mu)^{\dot\al}
+i\th \si ^\mu\bar\ep \prt _\mu
-i\ep \si ^\mu\bar\th \prt _\mu\right ]\Ph_k(x,\th,\bar\th)\;,
\label{susyCPT}
\eea
closes by 
construction on real superfields, ${\cal S}^*={\cal S}$. However, it 
{\it does not} close on the superfields $\Ph_k$ and $\Ph^*_k$
as can be explicitly checked by 
a short calculation\footnote{Application of the supersymmetric 
transformation in Eqn.~\rf{susyCPT} to the chiral superfield $\Ph_k$ 
generates components in the antichiral superfield $\Ph^*_k$, and vice 
versa.}. The obstruction to defining a 
supersymmetry generator on chiral superfields was already pointed out in 
Ref.~\cite{Berger:2001rm}. 
The objects $\Ph_k$ and $\Ph^*_k$ are conjugates and can be used to construct
real superfields. 
Since the supersymmetric transformation 
in Eqn.~\rf{susyCPT} closes on all real superfields, it closes on 
$\Ph_k^*\Ph_k$.

\section{Superspace Transformations}

Now we proceed to show how the Lorentz-violating extensions of the Wess-Zumino
model can be understood as transformations on the superfields.
Define
\bea
&&X\equiv (\th \sigma ^\mu \bar\th)\prt _\mu\;, \\
&&Y\equiv k_{\mu\nu}(\th \sigma ^\mu \bar\th)\prt ^\nu\;, \\
&&K\equiv k_\mu(\th \sigma ^\mu \bar\th)\;,
\eea
so that
\bea
&&U_x \equiv e^{iX}=1+i(\th \sigma ^\mu \bar\th)\prt _\mu -{1\over 4}(\th \th)
(\bar\th \bar\th)\Box \;, \\
&&U_y \equiv e^{iY}=1+ik_{\mu\nu}(\th \sigma ^\mu \bar\th)\prt^\nu
 -{1\over 4}k_{\mu \nu}k^{\mu \rho}(\th \th)
(\bar\th \bar\th)\prt ^\nu \prt _\rho\;, \\
&&T_k \equiv e^{-K}=1-k_\mu(\th \sigma ^\mu \bar\th)+{k^2\over 4}(\th \th)
(\bar\th \bar\th)\;.
\eea
Since $X$ and $Y$ are derivative operators, the action of $U_x$ and $U_y$
on a superfield ${\cal S}$
can be understood as a coordinate shift. In the customary (Lorentz symmetric)
case involving 
$U_x$ one has
\bea
&&U_x{\cal S}(x,\th,\bar\th)={\cal S}(y,\th,\bar\th)\;,
\eea
i.e. the spacetime coordinate $x^\mu$ is shifted to $y^\mu$.
The chiral superfield $\Ph(x,\th,\bar\th)$ is a function of $y^\mu$ and 
$\th$ only, so it must then be of the form $\Ph(x,\th,\bar\th)=U_x\Psi(x,\th)$
for some function $\Psi$. The chiral superfield $\Ph(x,\th,\bar\th)$ 
does not depend on $\bar\th$ except through 
the coordinate $y^\mu$. As is well-known, the kinetic terms of the 
Wess-Zumino model can be expressed as 
\bea
&&\int d^4\th \left [U_x^*\Psi(x,\bar\th)^*\right ]\left [U_x\Psi(x,\th)
\right ]
=\int d^4\th \Ph^*(z,\bar\th)\Ph(y,\th)\;.
\eea

The supersymmetric models with Lorentz-violating terms 
can be expressed in terms of superfields in an
analogous way. Consider the superfields
\bea
\Ph_y(x,\th,\bar\th)&=&U_yU_x\Psi(x,\th)\;, \\
\Ph^*_y(x,\th,\bar\th) &=& U_y^*U_x^*\Psi^*(x,\bar\th)\nonumber \\ 
&=&U_y^{-1}U_x^{-1}\Psi^*(x,\bar\th)\;.
\eea
Applying $U_y$ to the chiral and antichiral superfields merely effects the
substitution $\prt _\mu \to \prt _\mu +k_{\mu\nu}\prt ^\nu$. Since $U_y$ 
involves a derivative operator just as $U_x$, the derivation of the chiral 
superfield $\Ph_y$ can be understood as a function of the variables $x_+^\mu$
and $\theta$ analogous to how, in the usual case, $\Ph$ is a function 
of the variables $y^\mu$ and $\theta$. The 
Lagrangian is given by 
\bea
&&\int d^4\th \Ph_y^*\Ph_y+ \int d^2\th \left [ 
{1\over 2}m\Ph_y^2 +{1\over 3}g\Ph_y^3 
+h.c.\right ]\nonumber \\
&&=\int d^4\th \left [U_y^*\Ph^*\right ]
\left [U_y\Ph\right ]+ \int d^2\th \left [ 
{1\over 2}m\Ph^2 +{1\over 3}g\Ph^3 
+h.c.\right ]\;.
\label{superspace2}
\eea

For the CPT-violating model the superfields have the form
\bea
\Ph_k(x,\th,\bar\th)&=&T_kU_x\Psi(x,\th)\;, \\
\Ph^*_k(x,\th,\bar\th) &=& T_k^*U_x^*\Psi^*(x,\bar\th)\nonumber \\ 
&=&T_kU_x^{-1}\Psi^*(x,\bar\th)\;.
\eea
It is helpful to note that the transformation $U_x$ acts on $\Psi$ and 
its inverse $U_x^{-1}$ 
acts on $\Psi^*$, while the same transformation $T_k$ acts on both 
$\Psi$ and $\Psi^*$ (rather than its inverse). A consequence of this fact is 
that the supersymmetry transformation will act differently on the components 
of the chiral superfield and its conjugate as described in 
Ref.~\cite{Berger:2001rm}. Specifically the chiral superfield $\Ph_k$ is the
same as $\Ph$ with the substitution $\prt _\mu \to \prt _\mu +ik_\mu$ 
whereas the 
antichiral superfield $\Ph_k^*$ is the same as $\Ph^*$ with the 
substitution $\prt _\mu \to \prt _\mu -ik_\mu$.

The CPT-violating model in Eqn.~\rf{CPT} 
can then be represented in the following way as a 
superspace integral:
\bea
&&\int d^4\th \Ph_k^*\Ph_k=\int d^4\th \Ph^* e^{-2K}\Ph
\label{proj}
\eea
The projection factor $e^{-2K}$ commutes through the superfields, but its 
placement in Eqn.~\rf{proj} is suggestive of the coupling of a chiral 
superfield to a gauge field\footnote{The kinetic terms of a 
supersymmetric gauge 
theory for which the kinetic terms can be expressed as 
\bea
&&\int d^4\th \Ph^* e^{2gV}\Ph 
\eea
where $V$ is a vector superfield, and the ordinary derivatives $\prt _\mu$ are
replaced by {\it gauge} covariant derivative $D_\mu =\prt _\mu \pm igv_\mu$
where $g$ is the gauge coupling and $v_\mu$ is the component of $V$ multiplying
$\th \si ^\mu \bar\th$. The sign in the covariant derivative is different for 
the chiral and antichiral superfields.}.
Unlike the CPT-conserving model, the $(\th \th)(\bar\th \bar\th)$ component
of $\Ph^*\Ph$ no longer transforms into a total derivative. A certain 
combination of components of $\Ph^*\Ph$ does transform into a total 
derivative, and this combination can be understood as being the 
$(\th \th)(\bar\th \bar\th)$ component of $\Ph_k^*\Ph_k$.
Therefore, we have achieved a superspace formulation of the 
CPT-violating supersymmetric model that was first described
in Ref.~\cite{Berger:2001rm}.

As mentioned above, the supersymmetry transformation does not 
close for a chiral superfield.
The components 
of the standard chiral supermultiplet $\Ph$ and its conjugate $\Ph^*$ each
transform into themselves under a CPT-transformation (as opposed to a parity 
transformation which interchange the components of the 
supermultiplet with those of its conjugate). 
On the other hand, for $\Ph_k$ and $\Ph_k^*$, the
CPT transformation takes $k_\mu \to -k_\mu$, so that the chiral and antichiral 
superfields $\Ph_k$ and $\Ph_k^*$ mix under it. 

\section{Conclusions}

Extensions of the Wess-Zumino model that contain terms that violate the 
Lorentz symmetry but preserve the supersymmetric part of the 
superPoincar\'e algebra exist. The simplest extension preserves CPT and is 
obtained by the substitution $\prt _\mu \to \prt _\mu +k_{\mu \nu}\prt ^\nu$.
Since this substitution replaces a derivative operator with another derivative
operator, the $(\th \th)(\bar\th \bar\th)$ component of a vector superfield
and the $\th \th$ component of functions of a  
chiral superfield still transform as total 
derivatives. The projection from superfields to components proceeds in the 
usual way. In fact one can introduce the new coordinate
$x_+^\mu =x^\mu +i\th \si ^\mu \bar \th +ik^{\mu \nu}\th \si _\nu \bar \th$ 
and obtain the chiral superfield as the most general function of 
this $y$ and $\th$. It is clear that adding Lorentz-violation in this 
fashion can be 
immediately extended to encompass supersymmetric gauge theories as well.

The CPT-violating model, however, does not involve adding a derivative 
operator. A certain combination of the components of the 
$\Ph ^* \Ph$ superfield does in fact transform into a total derivative.
This combination can be projected out of the vector superfield by 
applying the operator $e^{-2k_\mu(\th\si^\mu\bar\th)}$ and then 
performing the usual projection of the $(\th \th)(\bar\th \bar\th)$ component.
One obtains precisely the CPT-violating model presented 
in Ref.~\cite{Berger:2001rm} 

The conventional Wess-Zumino model can be described in 
terms of superspace transformations and projecting out the highest 
component of the result. 
It was shown that the two Lorentz-violating models can be understood 
in terms of similar transformations on the 
superfields.


\vspace{0.5cm}

\section*{Acknowledgments}

This work was supported in part by the U.S.
Department of Energy under Grant No. No.~DE-FG02-91ER40661.



\end{document}